\documentclass[showpacs,prl,onecolumn,aps,superscriptaddress,preprintnumbers,nofootinbib]{revtex4}
\usepackage[T1]{fontenc}
\usepackage[latin9]{inputenc}
\usepackage{amsmath,amssymb}
\usepackage{epsfig}
\usepackage{graphicx}
\usepackage{amsmath}
\usepackage{amsfonts}
\def\slashchar#1{\setbox0=\hbox{$#1$}     		
   \dimen0=\wd0                                 	
   \setbox1=\hbox{/} \dimen1=\wd1               	
   \ifdim\dimen0>\dimen1                        	
      \rlap{\hbox to \dimen0{\hfil/\hfil}}      	
      #1                                        	
   \else                                        	
      \rlap{\hbox to \dimen1{\hfil$#1$\hfil}}   	
      /                                         	
   \fi}

\renewcommand{\vec}{\boldsymbol}
\newcommand{\beq}{\begin{equation}}
\newcommand{\eeq}{\end{equation}}
\newcommand{\bea}{\begin{eqnarray}}
\newcommand{\eea}{\end{eqnarray}}
\newcommand{\baa}{\begin{array}}
\newcommand{\eaa}{\end{array}}

\def\eq#1{{Eq.~(\ref{#1})}}

\def\fig#1{{Fig.~\ref{#1}}}

\newcommand{\bas}{\bar{\alpha}_S}

\newcommand{\nn}{\nonumber}

\newcommand{\h}{\frac{1}{2}}

\newcommand{\Lb}{\left(}
\newcommand{\Rb}{\right)}

\renewcommand{\vec}[1]{\boldsymbol{#1}}

\begin{document}
\title{QCD Odderon: non linear evolution in the leading twist}
\author{Carlos Contreras}
\email{carlos.contreras@usm.cl}
\affiliation{Departamento de F\'isica, Universidad T\'ecnica Federico Santa Mar\'ia,  Avda. Espa\~na 1680, Casilla 110-V, Valpara\'iso, Chile}
\author{ Eugene ~ Levin}
\email{leving@tauex.tau.ac.il, eugeny.levin@usm.cl}
\affiliation{Departamento de F\'isica, Universidad T\'ecnica Federico Santa Mar\'ia,  Avda. Espa\~na 1680, Casilla 110-V, Valpara\'iso, Chile}
\affiliation{Centro Cient\'ifico-
Tecnol\'ogico de Valpara\'iso, Avda. Espa\~na 1680, Casilla 110-V, Valpara\'iso, Chile}
\affiliation{Department of Particle Physics, School of Physics and Astronomy,
Raymond and Beverly Sackler
 Faculty of Exact Science, Tel Aviv University, Tel Aviv, 69978, Israel}
\author{Rodrigo Meneses}
\email{rodrigo.meneses@uv.cl}
\affiliation{Escuela de Ingenier\'\i a Civil, Facultad de Ingenier\'\i a, Universidad de Valpara\'\i so, General Cruz 222, Valpara\'\i so, Chile}
\author{Michael Sanhueza}
\email{michael.sanhueza.roa@gmail.com}
\affiliation{Departamento de F\'isica, Universidad T\'ecnica Federico Santa Mar\'ia,   Avda. Espa\~na 1680, Casilla 110-V, Valpara\'iso, Chile}
\date{\today}

\begin{abstract}
In the paper we propose and solve  analytically the non-linear evolution equation in the
 leading twist approximation for the Odderon contribution. We found three
 qualitative features of this solution, which differs the Odderon
 contribution from the Pomeron one :(i) the behaviour in the vicinity of the saturation scale cannot be  derived from the linear evolution in  a dramatic difference with the Pomeron case;  (ii) a 
substantial
 decrease of the Odderon contribution with the energy; and (iii)    the
 lack of  geometric scaling behaviour. The two last   features have been seen  in numerical attempts to solve the Odderon equation.

   \end{abstract}

\maketitle

\vspace{-0.5cm}
\tableofcontents






\section{ Introduction}

The new data of the TOTEM collaboration\cite{TOTEMRHO1,TOTEMRHO2,TOTEMRHO3,TOTEMRHO4}
triggered  hot discussions on the Odderon: a state with negative signature
 and with an intercept, which is close to unity (see 
Refs.\cite{KMRO,BJRS,TT,MN,BLM,SS,KMRO1,KMRO2,GLP,CNPSS}).
 This state arises  naturally  in  perturbative QCD (see Ref.\cite{KOLEB}
 for the review). In Refs.\cite{BLV,KS} the linear equation for the
 perturbative Odderon has been derived  and it has been shown,  that
 the intercept of the Odderon  is equal $\alpha_{\rm Odd}\Lb t=0\Rb\,\,
=\,\,1$.  Having negative signature such  an Odderon generates the real 
part
 of the scattering amplitude, which does not depend on energy.
  Specifically such an Odderon has been discussed in the phenomenological 
attempts
 to
 describe the experimental data in Refs.
\cite{KMRO,BJRS,TT,MN,BLM,SS,KMRO1,KMRO2,GLP,CNPSS}.

However, in the Colour Glass Condensate(CGC) approach, the energy 
dependence
 of the Odderon contribution is affected by the shadowing
 corrections\cite{KS,KOLEB}, which result  in a decrease of
 the Odderon amplitude with increasing energy. In this paper,
 we wish to discuss the  non-linear evolution of the Odderon
 in the CGC approach continuing research started by
 Refs.\cite{KS,KOLEB}.

We wish to recall that 
  CGC  approach is the only candidate for the effective theory at high
 energies, which is based on our microscopic theory: QCD
 (see Ref.\cite{KOLEB} for a review).  It has been shown
 that the non linear equations for the positive signature
 (Balitsky-Kovchegov (BK) equations\cite{BK}) take a simple
 form, if we use the simplified BFKL kernel\cite{BFKL} and restrict
  ourselves to the contribution of the leading twist only. 
In this paper we generalize this approach for the case of the
 Odderon contribution.

\section{  Balitsky-Kovchegov(BK) equation in the leading twist approximation}
The BK evolution equation  for the dipole-target scattering amplitude
 $N\Lb \vec{x}_{10},\vec{b},Y ; R\Rb$ has the general form in the leading
 order (LO)  of perturbative QCD ($R$ denotes the size of the
 target)\cite{KOLEB,BK,GLR,MUQI,MV}:
\bea \label{BK}
&&\frac{\partial}{\partial Y}N\Lb \vec{x}_{10}, \vec{b} ,  Y; R \Rb = \\
&&\bas\!\! \int \frac{d^2 \vec{x}_2}{2\,\pi}\,K\Lb \vec{x}_{02}, \vec{x}_{12}; \vec{x}_{10}\Rb \Bigg(N\Lb \vec{x}_{12},\vec{b} - \h \vec{x}_{20}, Y; R\Rb + 
N\Lb \vec{x}_{20},\vec{b} - \h \vec{x}_{12}, Y; R\Rb - N\Lb \vec{x}_{10},\vec{b},Y; R \Rb\nn\\
&&-\,\, N\Lb \vec{x}_{12},\vec{b} - \h \vec{x}_{20}, Y; R\Rb\,N\Lb \vec{x}_{20},\vec{b} - \h \vec{x}_{12}, Y; R\Rb\Bigg)
\eea
where $\vec{x}_{i k}\,\,=\,\,\vec{x}_i \,-\,\vec{x}_k$  and
 $ \vec{x}_{10} \equiv\,\vec{r}$, $\vec{x}_{20}\,\equiv\,\vec{r}' $ and $\vec{x}_{12} \,\equiv\,\vec{r}\,-\,\vec{r}'$.  $Y$ is the rapidity of the scattering dipole and $\vec{b}$ is the impact factor. $K\Lb \vec{x}_{02}, \vec{x}_{12}; \vec{x}_{10}\Rb$ is the kernel of the BFKL equation which in the leading order has the following form:
\beq \label{KERLO}
K^{\rm LO} \Lb \vec{x}_{02}, \vec{x}_{12}; \vec{x}_{10}\Rb\,\,=\,\,\frac{x^2_{10}}{x^2_{02}\,x^2_{12}}
\eeq

 For the kernel of the LO BFKL equation (see \eq{KERLO}) the eigenvalues
  take the form\cite{BFKL,LIP}:

\beq \label{CHI}
\omega\Lb \bas, \gamma\Rb\,\,=\,\,\bas\,\chi^{LO}\Lb \gamma \Rb\,\,\,=\,\,\,\bas \Lb 2 \psi\Lb 1\Rb \,-\,\psi\Lb \gamma\Rb\,-\,\psi\Lb 1 - \gamma\Rb\Rb
\eeq
where $\psi(z)$  is the Euler psi-function $\psi\Lb z\Rb =
 d \ln \Gamma(z)/d z$.
The general BFKL kernel of \eq{KERLO} and \eq{CHI} has 
 contributions of all possible twists, and it cannot be solved
 analytically. However, as it was shown in Ref.\cite{LETU} the
 situation becomes much simpler if we restrict ourselves to the
 leading twist contribution to the BFKL kernel, which has the
 form\cite{LETU} 
\bea \label{SIMKER}
\chi\Lb \gamma\Rb\,\,=\,\, \left\{\begin{array}{l}\,\,\,\frac{1}{\gamma}\,\,\,\,\,\,\,\,\,\,\mbox{for}\,\,\,\tau\,=\,r Q_s\,>\,1\,\,\,\,\,\,\mbox{summing} \Lb \ln\Lb r Q_s\Rb\Rb^n;\\ \\
\,\,\,\frac{1}{1 \,-\,\gamma}\,\,\,\,\,\mbox{for}\,\,\,\tau\,=\,r Q_s\,<\,1\,\,\,\,\,\mbox{summing}
\Lb \ln\Lb1/(r\,\Lambda_{\rm QCD})\Rb\Rb^n;\\  \end{array}
\right.
\eea
instead of the full expression of \eq{CHI}.

In the saturation region 
 where $\tau\,\,>\,\,1$ the logs 
   originate from the decay of a large size dipole into one small
 size dipole  and one large size dipole\cite{LETU}.  However, the size of the
 small dipole is still larger than $1/Q_s$. This observation can be
 translated to the following form of the kernel in the LO
\bea \label{K2}
\frac{\bas}{2 \pi}\int \, \displaystyle{K\Lb \vec{x}_{01};\vec{x}_{02},\vec{x}_{12}\Rb}\,d^2 x_{02} \,&\rightarrow&
\,\frac{\bas}{2}\, \int^{x^2_{01}}_{1/Q^2_s(Y,b)} \frac{ d x^2_{02}}{x_{02}^2}\,\,+\,\,
\frac{\bas}{2}\, \int^{x^2_{01}}_{1/Q^2_s(Y, b)} \frac{ d |\vec{x}_{01}  -
 \vec{x}_{02}|^2}{|\vec{x}_{01}  - \vec{x}_{02}|^2}\,\,\nn\\
 &=&\,\,\frac{\bas}{2}\, \int^{\xi}_{\xi_s} d \xi_{02} \,\,+\,\,\frac{\bas}{2}\, \int^{\xi}_{\xi_s} d \xi_{12}\eea
 where $\xi_{ik} \,=\,\ln\Lb x^2_{ik} Q_s^2(Y=Y_0)\Rb$.

Inside the saturation region the BK equation in  LO,  takes the form
\beq \label{BK2}
\frac{\partial^2 \widetilde{N}\Lb Y; \xi, \vec{b}\Rb}
{ \partial Y\,\partial \xi}\,\,=\,\, \bas \,\left\{ \Lb 1 
\,\,-\,\frac{\partial \widetilde{N}\Lb Y; \xi, \vec{b}
 \Rb}{\partial  \xi}\Rb \, \widetilde{N}\Lb Y; \xi,
 \vec{b}\Rb\right\}
\eeq
where 
 $\widetilde{N}\Lb Y; \xi, \vec{b}\Rb\,\,=\,\,\int^{\xi}_{\ln Q^2_s(Y)} d \xi'\,N\Lb Y; \xi', \vec{b}\Rb$ .
    
Introducing
\beq \label{BK3}
N\Lb Y; \xi', \vec{b}\Rb\,\,=\,\,1\,\,-\,\,\Delta\Lb  Y; \xi', \vec{b}\Rb\,\,=\,\,1 \,\,-\,\,e^{ - \Omega\Lb  Y; \xi', \vec{b}\Rb}
\eeq
we can reduce \eq{BK2} to the following expressions:
\beq \label{BK4}
\frac{ \partial \,\Omega\Lb  Y; \xi', \vec{b}\Rb}{\partial\,Y}\,\,=\,\,\bas \,\,\widetilde{N}\Lb Y; \xi, \vec{b}\Rb;~~~~~~~~
\frac{ \partial^2 \,\Omega\Lb  Y; \xi', \vec{b}\Rb}{\partial\,Y\,\,\partial\,\xi}\,\,=\,\,\bas\,\Bigg( 1\,\,-\,\,e^{-\,\Omega\Lb  Y; \xi', \vec{b}\Rb}\Bigg)
\eeq
Looking for the traveling wave solution
 (geometric scaling\cite{BALE,MUT,IIML,SGBK}) , we assume 
that $\Omega\Lb  Y; \xi', \vec{b}\Rb\,\,\equiv\,\,\Omega\Lb  z\Rb$ with 
\beq \label{Z}
z\,\,=\,\,\ln \tau\,\,=\,\,\ln\Lb Q^2_s\Lb Y, b\Rb\,r^2\Rb\,\,=\,\,4\,\bas\,Y\,\,+\,\,\xi
\eeq
\eq{BK4} takes the form:
\beq \label{BK5}
\frac{ d^2 \,\Omega\Lb  z \Rb}{d\,z^2}\,\,=\,\,\frac{1}{4}\,\Bigg( 1\,\,-\,\,e^{-\,\Omega\Lb  z\Rb}\Bigg)
\eeq
which has the solution (see formula {\bf 3.4.1.1} of  Ref.\cite{MATH}):
\begin{equation} \label{SOL}
\sqrt{2}\int^{\Omega}_{\Omega_0} \,d \Omega'\,\,\frac{  1}{ \sqrt{\Omega' \,\,+\,\,e^{- \Omega'} \,\,-\,\,1\,\,+\,\,{\rm C}}}\,\,=\,\,z
\end{equation}
for the function $\Omega$.

The value of $C$ has to be determined from matching with the region
 $\tau \,<\,1$. For small $\Omega_0$ \,\,$C=0$. Indeed, in this
 case the solution at small $\Omega$   has the following form: 
\begin{equation} \label{SOLSPHI}
\Omega\,\,=\,\,\Omega_0 \,e^{\frac{1}{2} \,z}
\end{equation}
which coincides with the general solution\cite{MUT} 
 for the region $\tau\,<\,1$ at
 small $\Omega_0 \,=\,N_0\,\ll\,1$.
\section{Linear evolution in perturbative QCD region}
\subsection{The BFKL equation}
\begin{figure}
\centering
      \includegraphics[width=6cm]{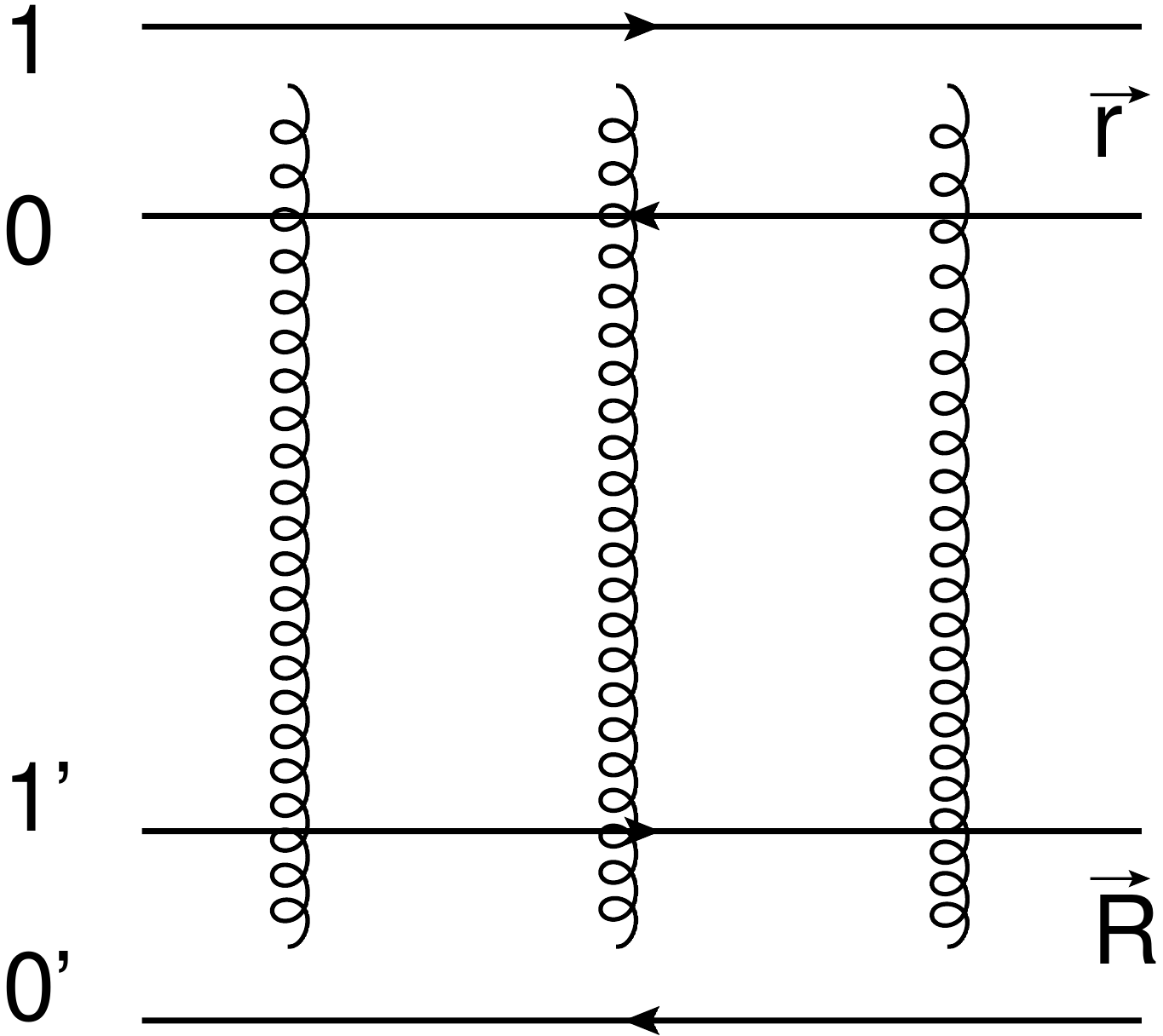} 
         \caption{The initial condition for evolution of the Odderon: the
  three gluon exchange in Born Approximation of perturbative QCD.     }      
\label{ba}
\end{figure}

The linear  equation for the Odderon is the same as the BFKL equation,
 which takes the form:

\beq \label{BFKL}
\hspace{-0.2cm}\frac{\partial}{\partial Y}O\Lb \vec{x}_{10},\vec{b} ,  Y; R \Rb = 
\bas\!\! \int \frac{d^2 \vec{x}_2}{2\,\pi}\,K\Lb \vec{x}_{02}, \vec{x}_{12}; \vec{x}_{10}\Rb \Bigg\{O\Lb \vec{x}_{12},\vec{b} - \h \vec{x}_{20}, Y; R\Rb + 
O\Lb \vec{x}_{20},\vec{b} - \h \vec{x}_{12}, Y; R\Rb - O\Lb \vec{x}_{10},\vec{b},Y; R \Rb\Bigg\}
\eeq
Therefore, the difference between the BFKL Pomeron and  Odderon stems from
 the initial conditions and the signature: positive for the BFKL Pomeron
 and negative for the Odderon. As it is shown in Ref.\cite{LIP,KS} the
 general solution to \eq{BFKL} can be written as
\beq \label{SOL1}
O\Lb \vec{x}_{10},\vec{b} ,  \vec{x}_{1'0'}; Y \Rb\,\,=\,\,c_0\bas^3\frac{6}{\pi^3}\sum^{\infty}_{k=0}\int^{ \infty + i \epsilon}_{-\infty + i \epsilon}\!\!\!\! \!\!d \nu\, e^{ \omega\Lb \bas, k, \nu\Rb\,Y} \,\chi\Lb k, \nu\Rb\,\frac{ \nu^2 + \frac{(2 k + 1)^2}{4}}{(\nu^2 + k^2)\,(\nu^2\,+\,(k + 1)^2)}G_{\nu, k}\Lb \vec{\rho}_1,\vec{\rho}_0,\vec{\rho}_{1'}, \vec{\rho}_{0'}\Rb
\eeq
where the eigenvalues $\omega\Lb \bas, k, \nu\Rb$ are equal to \cite{LIP}
\bea \label{SOL2}
\omega\Lb \bas, k, \nu\Rb\,\,\,=\,\,\,\bas\,\chi\Lb k, \nu\Rb\,\,\,&=&\bas\Big( 2\,\psi(1)\,-\,\psi( \frac{1+|n|}{2}\, +\, i\nu) \,-\,\psi(\frac{1+|n|}{2} \,-\,i\nu)\Big)\nn\\
\Lb |n|\,=\,2\,k\,+\,1\Rb&=&\,\,\,\bas\Big( 2\,\psi(1)\,-\,\psi( k\, + \,1\, +\, i\nu) \,-\,\psi(k\, + \,1 \,-\,i\nu)\Big)
\eea
and the eigenfunctions  have the following form\cite{LIP,NAPE}:
\beq \label{EGNFUN}
G_{\nu, k}\Lb \vec{\rho}_1,\vec{\rho}_0,\vec{\rho}_{1'}, \vec{\rho}_{0'}\Rb\,\,=\,\,c_1\, w^h\,w^{*\tilde{h}}{}_2F_3\Lb h,h,2\,h,w\Rb\,\,+\,\,c_2 \,w^{1- h}\,w^{* 1 - \tilde{h}}{}_2F_3\Lb 1 - \tilde{h}, 1 - \tilde{h},2\,-2\,\tilde{h},w^*\Rb
\eeq
where ${}_2F_3$ denotes the hypergeometric function (see formula {\bf 9.1}
 in Ref.\cite{RY}),
\beq \label{W}
w\,\,=\,\,\frac{\rho_{01}\,\rho_{0' 1'}}{\rho_{0 0'}\,\rho_{1 1'}}; ~~~~\,\,h\,=\,k\,+\,1\,+\,i\,\nu; ~~~\,\,\,\tilde{h}\,=\,\,\,-\,k\,\,+\,\,i\,\nu;
\eeq
and
\beq \label{C}
c_2\,=\,\frac{b_{k, \nu}}{2\,\pi^2};~~~\frac{c_1}{c_2}\,=\,\frac{b_{k,- \nu}}{b_{k, \nu}};~~~b_{k,\nu}\,\,=\,\,\pi^3\,2^{4\,i\,\nu}\frac{\Gamma\Lb - i \nu\,+\,k+1\Rb\,\Gamma\Lb i\,\nu\,+ k \,+\,\h\Rb}{\Gamma\Lb i\,\nu\,+\,k \,+\,1\Rb\,\Gamma\Lb - i\,\nu\,+ k + \h\Rb}
\eeq
We characterize all 
 two dimensional vectors,  shown in \fig{ba},  by
 complex numbers, specifically
\beq \label{VEC}
\rho_k\,=\,x_k + i\,y_k; ~~~~\rho^*_{k}\,=\,=\,x_k - i\,y_k;~~~\vec{\rho}_k\,=\,(x_k, y_k);
\eeq

\eq{SOL1} differs from the solution for the BFKL Pomeron, since the sum in
 this equation is over odd   $n = 2\,k + 1$, while  in the case of the
 BFKL Pomeron we sum over even  $n\,=\, 2\,k$. Indeed, the Odderon
 corresponds to the negative signature which  generates the states
 that change  sign under  charge conjugation, which, as was
 shown in Ref.\cite{KS}  corresponds to replacing quark $\vec{x}_1$
 by anti-quark $\vec{x}_0$, or in other words $\vec{x}_{01} \,\to\,- 
\vec{x}_{01}$. Since eigenfunctions under this transformation have the
 following properties:

\beq \label{CC}
G_{\nu, k}\Lb \vec{x}_{01},\vec{x}_{0'1'},\vec{b} \Rb \,\,=\,\,\Lb - 1\Rb^n\,G_{\nu, k}\Lb -\, \vec{x}_{01},\vec{x}_{0'1'},\vec{b} \Rb
\eeq

we see that the Pomeron and Odderon correspond to summation over even and
 odd $n$, respectively.
\eq{SOL1} satisfies the initial condition, which is given by the Born
 Approximation diagram of \fig{ba}\cite{KS}:
\beq \label{IC}
O_0\Lb \vec{x}_0, \vec{x}_1;Y=0\Rb\,\,=\,\,c_0 \bas^3\,\ln^3\Lb w\,w^*\Rb;~~~~~~~~~
c_0\,\,=\,\,\frac{(N_c^2-4)\,(N^2_c - 1)}{4\,N^3_c}
\eeq
where $N_c$ denotes the number of colours.

\begin{figure}
\begin{tabular}{c  c  c}
      \includegraphics[width=7.5cm]{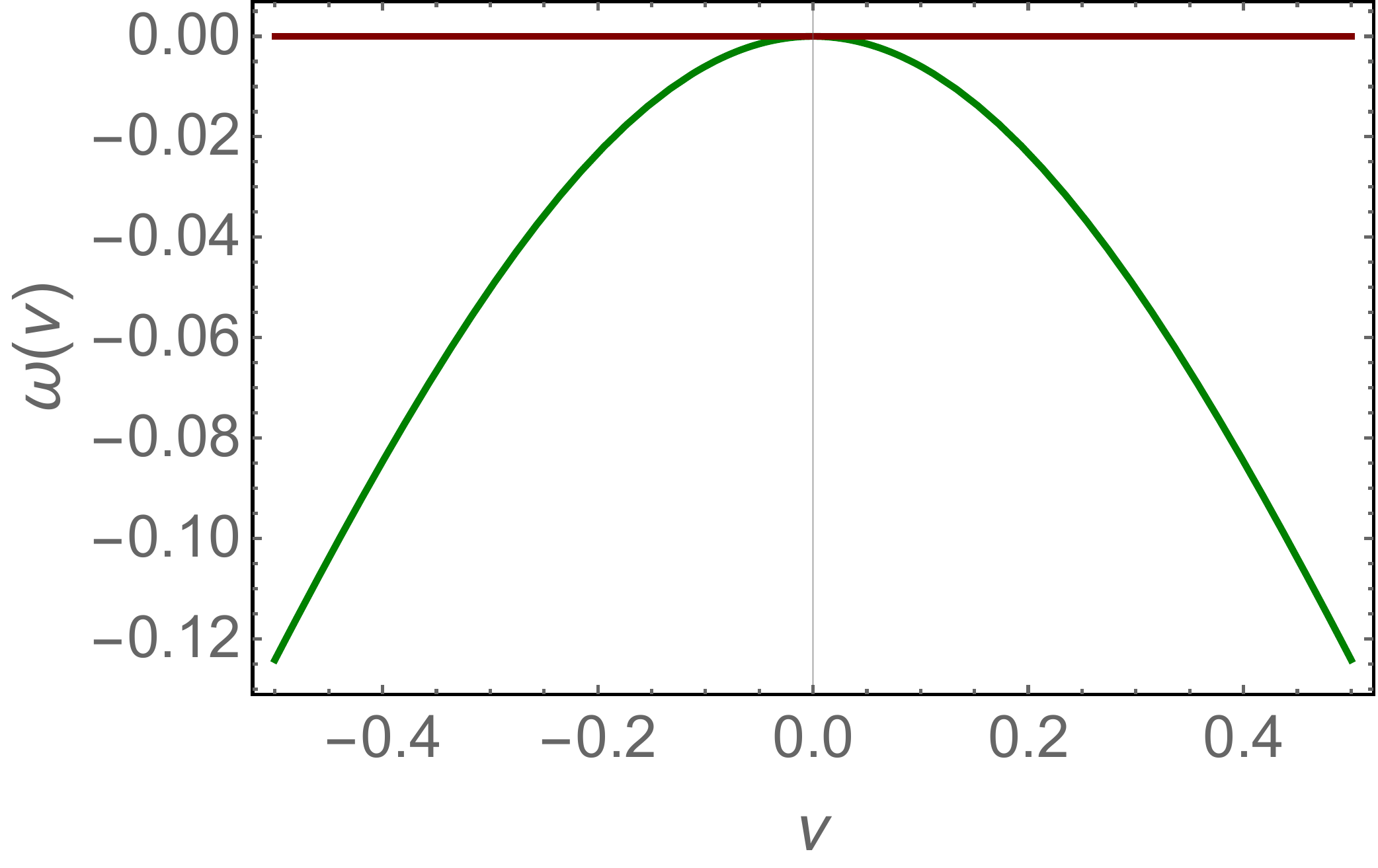}&  & \includegraphics[width=7cm]{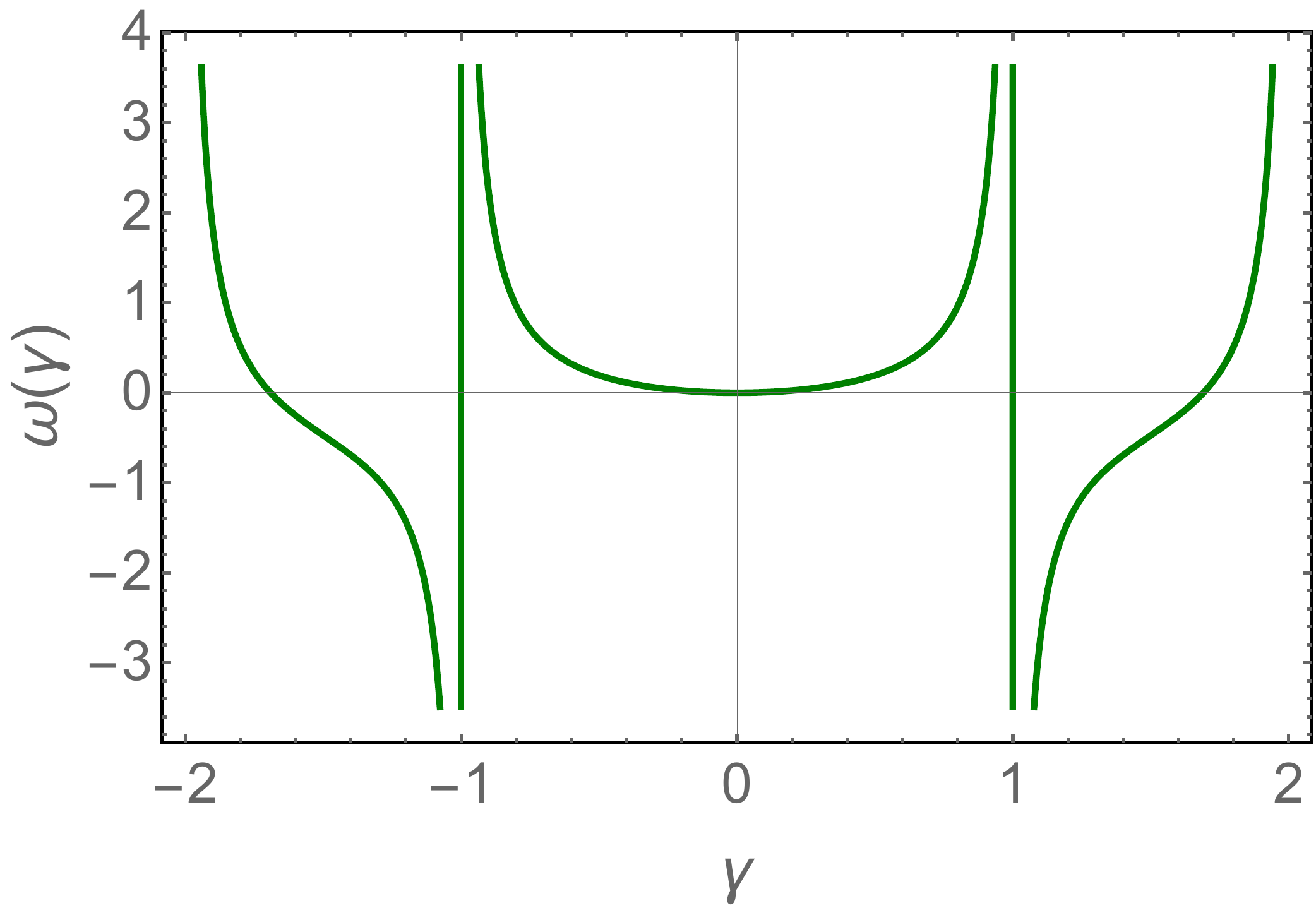}\\
      \fig{om}-a & ~~~~~~~~~~~&\fig{om}-b\\
 \end{tabular}
         \caption{\fig{om}-a: $\omega\Lb \bas,k=0,\nu\Rb$ versus $\nu$. The
 red line shows $\omega =0$.\fig{om}-b: The singular behaviour of  
 $\omega\Lb \bas,k=0,\nu\Rb$ versus $\gamma$, where $\gamma \,\equiv\,i\,\nu$.
     }      
\label{om}
\end{figure}

The main contribution in sum of \eq{SOL1}, stems from $k=0$. $\omega\Lb
 k=0,\nu\Rb$ is shown in \fig{om}-a. One can see that the maximal
 intercept is equal to 0. At $Y \,\gg\,Y_0$ the
 main contribution to \eq{SOL1}, gives the term with $k$=0 and,
 hence, the Odderon contribution
for $w \,w^*\,\ll\,1$ takes the form
:

\bea \label{SOL3} 
&&O\Lb \vec{x}_{10},\vec{b} ,  Y; \vec{x}_{1'0'}; Y \Rb\,\,=\\
&&\,\,c_0\bas^3\frac{6}{\pi^3}\,\int^{ \infty + i \epsilon}_{-\infty + i \epsilon}\!\!\!\! \!\!d \nu\, e^{ \omega\Lb \bas, 0, \nu\Rb\,Y} \,\chi\Lb 0, \nu\Rb\,\frac{ \nu^2 + \frac{1}{4}}{(\nu^2 )\,(\nu^2\,+\,1)}\Bigg\{ c_1\Lb  \frac{w}{w^*}\Rb^{\h} \Lb w\,w^*\Rb^{\h + i \,\nu} \,\,+\,\,c_2\Lb  \frac{w}{w^*}\Rb^{-\h} \Lb w\,w^*\Rb^{\h - i \,\nu}\Bigg\}\nn\\
&& = c_0\bas^3\frac{6}{\pi^3}\,\int^{ \infty + i \epsilon}_{-\infty + i \epsilon}\!\!\!\! \!\!d \nu\, e^{ \omega\Lb \bas, 0, \nu\Rb\,Y} \,\chi\Lb 0, \nu\Rb\,\frac{ \nu^2 + \frac{1}{4}}{(\nu^2 )\,(\nu^2\,+\,1)}\Bigg\{ c_1\Lb  \frac{w}{w^*}\Rb^{\h} e^{\Lb \h + i \,\nu\Rb\,\xi} \,\,+\,\,c_2\Lb  \frac{w}{w^*}\Rb^{-\h}e^{\Lb \h - i \,\nu\Rb\,\xi}\Bigg\}
\nn\eea
with
\beq \label{WW}
w\,w^*\,\,=\,\, \frac{ r^2\,R^2}{\Lb \vec{b}  + \h(\vec{r} - \vec{R})\Rb^2\,\Lb \vec{b}  -  \h(\vec{r} - \vec{R})\Rb^2}\,\,\equiv\,\,e^{\xi}
\eeq
where $b$ denotes the impact parameter of two colliding dipoles with
 sizes $r$ and $R$ (see \fig{ba}).

\subsection{Diffusion  approximation}


 We can evaluate the integral of \eq{SOL3} using the expansion
 of $\chi\Lb 0,\nu\Rb $ with respect to $\nu$. Specifically,
 
 \beq \label{SOL4}
 \chi\Lb k = 0,\nu\Rb\,\,=\,\,-\,D\,\nu^2\,\,=\,
 \,\,-\,2 \,\zeta\Lb 3\Rb\nu^2\,\,+\,\,{\cal O} \Lb\nu^3\Rb
 \eeq
 where $\zeta\Lb 3\Rb$ is the Riemann $\zeta$ function
 ($\zeta\Lb 3\Rb\,\approx\,1.2$)\cite{RY}.
 
 Using the method of steepest descent  we see that \eq{SOL4}
  leads to the following estimates for \eq{SOL3}:
 \beq \label{SOL5}
O\Lb \vec{x}_{10},\vec{b} ,  Y; \vec{x}_{1'0'}; Y \Rb\,\,=\,\,c_0\bas^3\frac{3}{\pi^{3/2}}\,\sqrt{\frac{D}{2\,Y}} e^{ - \frac{ \xi^2}{4\,\bas\,D\,\, Y}} \,\Lb  c_1\,w\,+\,c_2 \,w^*\Rb
\eeq
 Therefore, the $\xi$ and $Y$ dependence for the Odderon look similar
 to the BFKL Pomeron in the diffusion approximation(see Ref.\cite{KOLEB}),
  but the value of $D$ for the Odderon in 7 times smaller than for the 
Pomeron.


\subsection{Double Log Approximation(DLA )  in coordinate representation}


In \fig{om}-b we plot the dependence of the eigenvalues
 $\omega\Lb \bas,k=0,\gamma\Rb$ versus $\gamma \,=\,i\,\nu$.
  This eigenvalue has two singularities at
 $\gamma\,=\, \pm 1$. In the vicinities of these points
 $\omega\Lb  \bas,k=0,\gamma\Rb\,\,\propto \frac{\bas}{1 \pm \gamma}$.
 From past experience with the BFKL Pomeron we expect that such 
singularities
 generate the double log contributions.
We can see this explicitly by re-writing \eq{SOL3} in the new notation
 $$ \Lb w\,w^*\Rb^\gamma\, =\,
e^{ \gamma \xi}\, = \,e^{ -\gamma \xi'}=\,e^{\Lb 1 \,-\,\gamma\Rb\xi'\,\,-\,\,\xi'}$$
with $\xi'\,=\,- \xi\,\,=\,\,\ln\Lb\frac{1}{w\,w^*}\Rb\,>\,0$ for small
 dipole size  $r\,\,\ll\,\,R\,\leq\,b$. With the new variables the 
contribution
 of the first term of \eq{SOL3} takes the form:

\bea \label{DLA1}
O^{\rm DLA}\Lb \xi; Y \Rb\,\,&=&\,\,
\,\,c_0\bas^3\frac{12}{\pi^2}\,\int^{ \epsilon + i \infty }_{  \epsilon - i \infty}\!\!\!\! \frac{d \gamma}{2\,i\,\pi}\, e^{ \frac{\bas}{ 1 - \gamma} \,Y\,\,+\,\,\Lb 1 - \gamma\Rb\,\xi'} \,\Bigg\{\chi\Lb 0, \gamma\Rb\,\frac{ -\gamma^2 + \frac{1}{4}}{(-\gamma^2 )\,(-\gamma^2\,+\,1)} c_1(\gamma)\Bigg\}\Big( w  w\,w^*\Big)\nn\\
&=& \,\,c_0\bas^3\frac{12}{\pi^2}\,\int^{ \epsilon + i \infty }_{  \epsilon - i \infty}\!\!\!\! \frac{d \gamma}{2\,i\,\pi}\,\sum_{n=0}^{\infty} \Lb \frac{\bas Y}{1 - \gamma}\Rb^n e^{ \,\,\Lb 1 - \gamma\Rb\,\xi'} \,\Bigg\{\chi\Lb 0, \gamma\Rb\,\frac{ -\gamma^2 + \frac{1}{4}}{(-\gamma^2 )\,(-\gamma^2\,+\,1)} c_1(\gamma)\Bigg\}\Big( w  w\,w^*\Big)\nn\\
 &=& {\rm Const} \,\sum_{n=0}^\infty \frac{ \Lb \bas\,Y\,\xi'\Rb^n}{\Lb n! \Rb^2}\Big( w  w\,w^*\Big)\,\,\,\equiv\,\,\Big( w  w\,w^*\Big)\,o\Lb \bas \,Y\,\xi'\Rb 
 \eea
where in ${\rm Const}$ we absorbed all constant factors. Deriving the last
 equation we use that
$\Big\{\dots\Big\}\xrightarrow{\gamma \to 1}  \frac{1}{1-\gamma}$.
 It is easy to show that  the second term has the contribution which  can
 be written in the form: $\Big( w^*  w\,w^*\Big)\,o\Lb \bas \,Y\,\xi'\Rb$.

  Generally  the DLA contribution occurs in the form:
\beq \label{DLA2}
O^{\rm DLA}\Lb \xi; Y \Rb\,\,=\,\,\Lb w \,+\,w^*\Rb  w\,w^*\,o^{\rm DLA}\Lb \bas \,Y\,\xi'\Rb
\eeq

 The same structure  can  be seen directly from \eq{BFKL}. Indeed,
 the DL contribution stems from the sizes
 $x_{02} \,\sim\,x_{12}\,\gg\,x_{10}$. In this kinematic region the
 BFKL kernel has a simple form (see \eq{KERLO}
\beq \label{DLAKER}
K^{\rm LO}\Lb  \vec{x}_{02}, \vec{x}_{12}; \vec{x}_{10}\Rb\,\,\xrightarrow{x_{02} \,\sim\,x_{12}\,\gg\,x_{10}}\,\,x^2_{10} \Bigg\{ \frac{1}{x_{02}^4}\,\,+\,\,\frac{2 \vec{x}_{01} \cdot \vec{x}_{02}}{x_{02}^6}\Bigg\}\,\,=\,\,
x^2_{10} \Bigg\{ \frac{1}{x_{02}^4}\,\,+\,\, 2\,\frac{x_{10}}{x_{02}^5}\,\cos\varphi \Bigg\}
\eeq
where $\varphi$ denotes the angle between vectors $\vec{x}_{10}$ and 
$\vec{x}_{02}$.

Plugging this kernel in \eq{BFKL} we see that \eq{BFKL} takes the form:

\beq \label{DLA3}
\hspace{-0.2cm}\frac{\partial}{\partial Y}O^{\rm DLA}\Lb \xi' ,  Y \Rb = 
\bas\!\! \int \frac{d^2 \vec{x}_{02}}{2\,\pi}\,\,
x^2_{10} \Bigg\{ \frac{1}{x_{02}^4}\,\,+\,\,2\,\frac{x_{10}}{x_{02}^5}\,\cos\varphi \Bigg\}\,O^{\rm DLA}\Lb \xi''; Y \Rb
\eeq
Recalling that $ w\,w^* \,\,\propto\,\,x^2_{02}$ and that  \eq{DLA2}
 can be re-written as follows:
\beq \label{DLA30}
O^{\rm DLA}\Lb \xi"; Y \Rb\,\,=\,\,\cos\varphi  (w\,w^*)^{3/2} \,o^{\rm DLA}\Lb \bas \,Y\,\xi''\Rb
\eeq
one can see that the first term in \eq{DLA3} vanishes due to integration
 over $\varphi$ and the second term  can be re-written as
\beq \label{DLA4}
\frac{\partial}{\partial Y}\,o^{\rm DLA}\Lb \xi' ,  Y \Rb\,\, = \,\,
\bas\,\int^{\xi'}\!\!\! \!\!\! \,d \xi'' \,o^{\rm DLA}\Lb \xi''; Y \Rb
\eeq

\subsection{Geometric scaling behaviour of the scattering amplitude in the
 vicinity of the saturation momentum}


It is well known\cite{KOLEB,GLR,BALE,LETU, IIML,MUT}, that for finding
 the saturation momentum, as well as for discussing the 
behaviour of the scattering amplitude in the vicinity of the saturation
 scale, we do not need to know the precise structure of the non-linear
 corrections. What we need, is to find the solution of the
 linear BFKL equation, which is a wave package that satisfies
 the condition, that phase and group velocities are equal\cite{GLR}.
 In other words, we need to take the integral in \eq{SOL3} by the method 
of steepest descend, and to satisfy two conditions for the saddle point
 ($\nu_{\rm sp}$):

\beq \label{VSQ}
(1)~~\frac{d\,\omega\Lb \bas, 0, \nu_{\rm sp}\Rb}{d \nu_{\rm sp} }\,Y\,\,+\, i\,\xi\,\,=\,\,0;~~~~
(2)~~\omega\Lb \bas, 0, \nu_{\rm sp}\Rb\,Y\,\,+\,\,\Lb \h + i\,\nu_{\rm sp}\Rb\,\xi\,\,=\,\,0
\eeq

The first equation determines the trajectory of the wave package, while
 the second fixes the front line on which our wave function is a
 constant. Dividing one equation by the second one, we obtain the
 following equation, which actually gives  $v_{\rm group} \,=\,v_{\rm phase}$:
\beq \label{VSQ1}
\frac{d\,\omega\Lb \bas, 0, \nu_{\rm sp}\Rb}{d \nu_{\rm sp} }\,\,=\,\,v_{\rm group} \,\,=\,\,v_{\rm phase}\,\,=\,\,\,\,i\,\frac{\omega\Lb \bas, 0, \nu_{\rm sp}\Rb}{\h + i\,\nu_{\rm sp}}
\eeq
\begin{figure}
\centering
      \includegraphics[width=7cm]{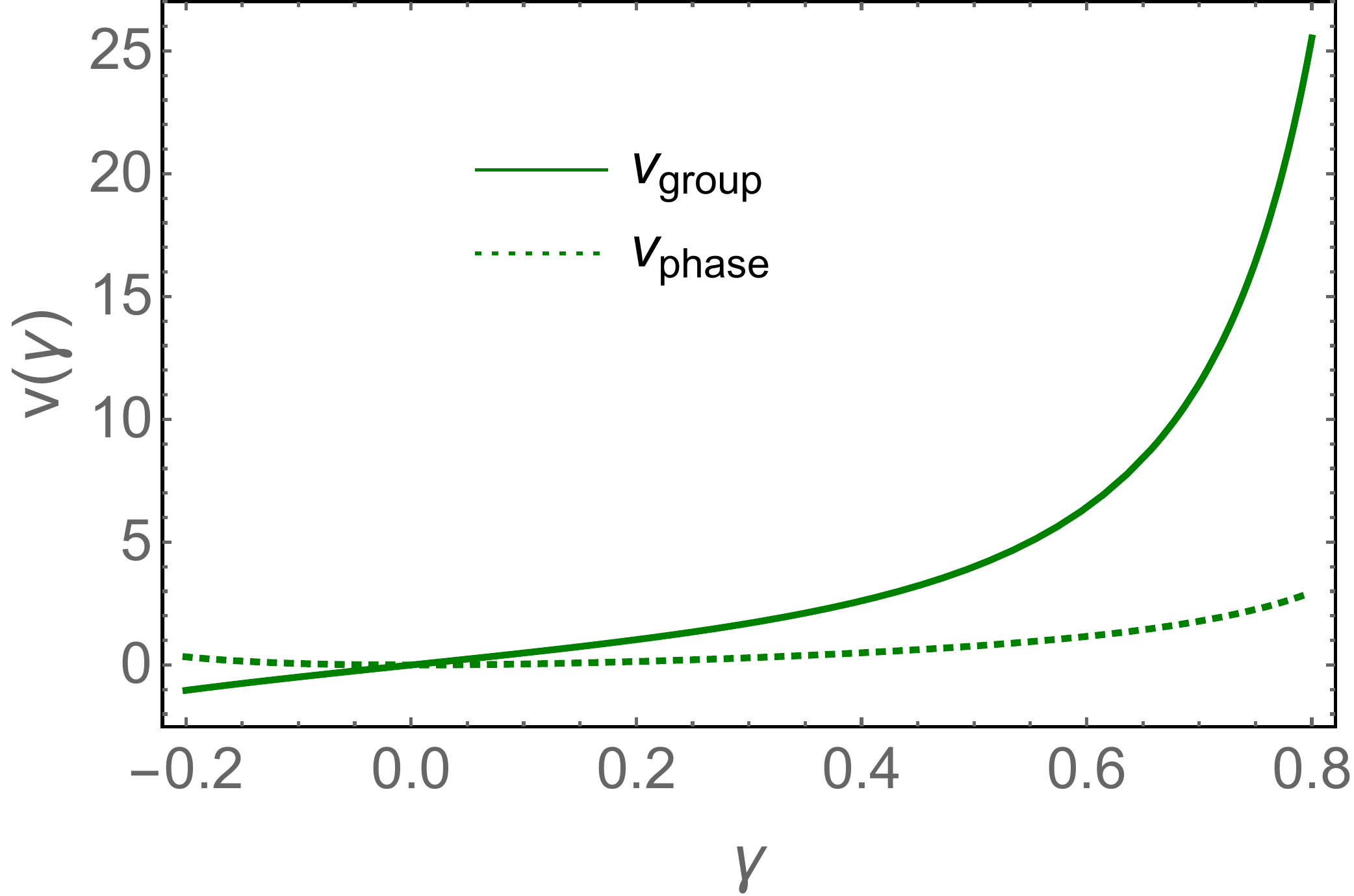} 
         \caption{ \eq{VSQ1}: $v_{\rm group}$ and $v_{\rm phase}$
 versus $ \gamma \equiv\,i\,\nu$.   }      
\label{gacr}
\end{figure}
In \fig{gacr} we plot the l.h.s. and the r.h.s. \eq{VSQ1}. One can see
 that this equation has no solution except at $\gamma=0$. However, at
 $\gamma \to 0$  $v_{\rm group} \,\propto\,\gamma^2$ 
while $v_{\rm phase} \,\propto\,\gamma$ and, therefore, \eq{VSQ1} does
 not have solution at small $\gamma$.

 This can be seen directly from \eq{SOL5} . At first sight   the diffusion
 solution of \eq{SOL5} is constant for 
$\xi$ determined from the following equation:
\beq \label{QS}
-\frac{\xi^2}{4\,\bas\,D\,Y} \,+\,\xi\,\,=\,\,0
\eeq
For $R\,\gg\,r,b$ $\xi =\,\ln\Lb r^2/R^2\Rb \,<\,0$. One can see that
 \eq{QS} has no solution for negative $\xi$. Recall, that in section
 III-B we found the saddle point, but without an additional condition
 of \eq{VSQ}-2.

Hence, the situation with the Odderon turns out to be quite different
 from the BFKL Pomeron: the linear equation does not provide 
 saturation, which might or might not stem from the solution to
 the non-linear equation, that was derived in Ref.\cite{KS} (see
 also Ref.\cite{KOLEB}). We also see no reason for the geometric
 scaling behaviour for the Odderon contribution. However, there is still
 the possibility that the non-linear evolution will lead to a
 geometric scaling solution, with the saturation scale determined by
 \eq{BK2}.


\section{Non-linear evolution for the Odderon}

\subsection{General equation}

The non-linear evolution equation for the  Odderon is derived
 in Ref.\cite{KS} (see also Ref.\cite{KOLEB}). It takes the form:
\bea \label{NEQ1}
&&\frac{\partial}{\partial Y}O\Lb \vec{x}_{10}, \vec{b} ,  Y; R \Rb = \\
&&\bas\!\! \int \frac{d^2 \vec{x}_2}{2\,\pi}\,K\Lb \vec{x}_{02}, \vec{x}_{12}; \vec{x}_{10}\Rb \Bigg\{O\Lb \vec{x}_{12},\vec{b} - \h \vec{x}_{20}, Y; R\Rb + 
O\Lb \vec{x}_{20},\vec{b} - \h \vec{x}_{12}, Y; R\Rb
-\,\, O\Lb \vec{x}_{01},\vec{b}, Y; R\Rb\,\nn\\
&&-\,\,N\Lb \vec{x}_{20},\vec{b} - \h \vec{x}_{12}, Y; R\Rb\,O\Lb \vec{x}_{12},\vec{b} - \h \vec{x}_{20}, Y; R\Rb\,\,-\,\,N\Lb \vec{x}_{12},\vec{b} - \h \vec{x}_{20}, Y; R\Rb\,O\Lb \vec{x}_{20},\vec{b} - \h \vec{x}_{12}, Y; R\Rb\Bigg\}\nn
\eea
 where the amplitude $N$ is the solution of the BK equation
 (see \eq{BK}) for the Pomeron. 
 \beq \label{NEQ2}
 N\Lb \vec{x}_{01},\vec{b}, Y; R\Rb \,\,=\,\,1\,\,-\,\,\Delta\Lb  \vec{x}_{01},\vec{b}, Y; R\Rb\,\,=\,\,
 1\,\,-\,\,\exp\Lb - \Omega\Lb \vec{x}_{01},\vec{b}, Y; R\Rb\Rb
 \eeq
 We can re-write \eq{NEQ1} in the form:
 \bea \label{NEQ3}
&&\frac{\partial}{\partial Y}O\Lb \vec{x}_{10}, \vec{b} ,  Y; R \Rb \,\,=\,\,
\bas\!\! \int \frac{d^2 \vec{x}_2}{2\,\pi}\,K\Lb \vec{x}_{02}, \vec{x}_{12}; \vec{x}_{10}\Rb \Bigg\{-\,\, O\Lb \vec{x}_{01},\vec{b}, Y; R\Rb\,\,\\
&&+\,\,\Delta\Lb \vec{x}_{20},\vec{b} - \h \vec{x}_{12}, Y; R\Rb\,O\Lb \vec{x}_{12},\vec{b} - \h \vec{x}_{20}, Y; R\Rb\,\,+\,\,\Delta\Lb \vec{x}_{12},\vec{b} - \h \vec{x}_{20}, Y; R\Rb\,O\Lb \vec{x}_{20},\vec{b} - \h \vec{x}_{12}, Y; R\Rb\Bigg\}\nn\eea

\begin{figure}
\centering
\begin{tabular}{c c c}
 \includegraphics[width=7cm]{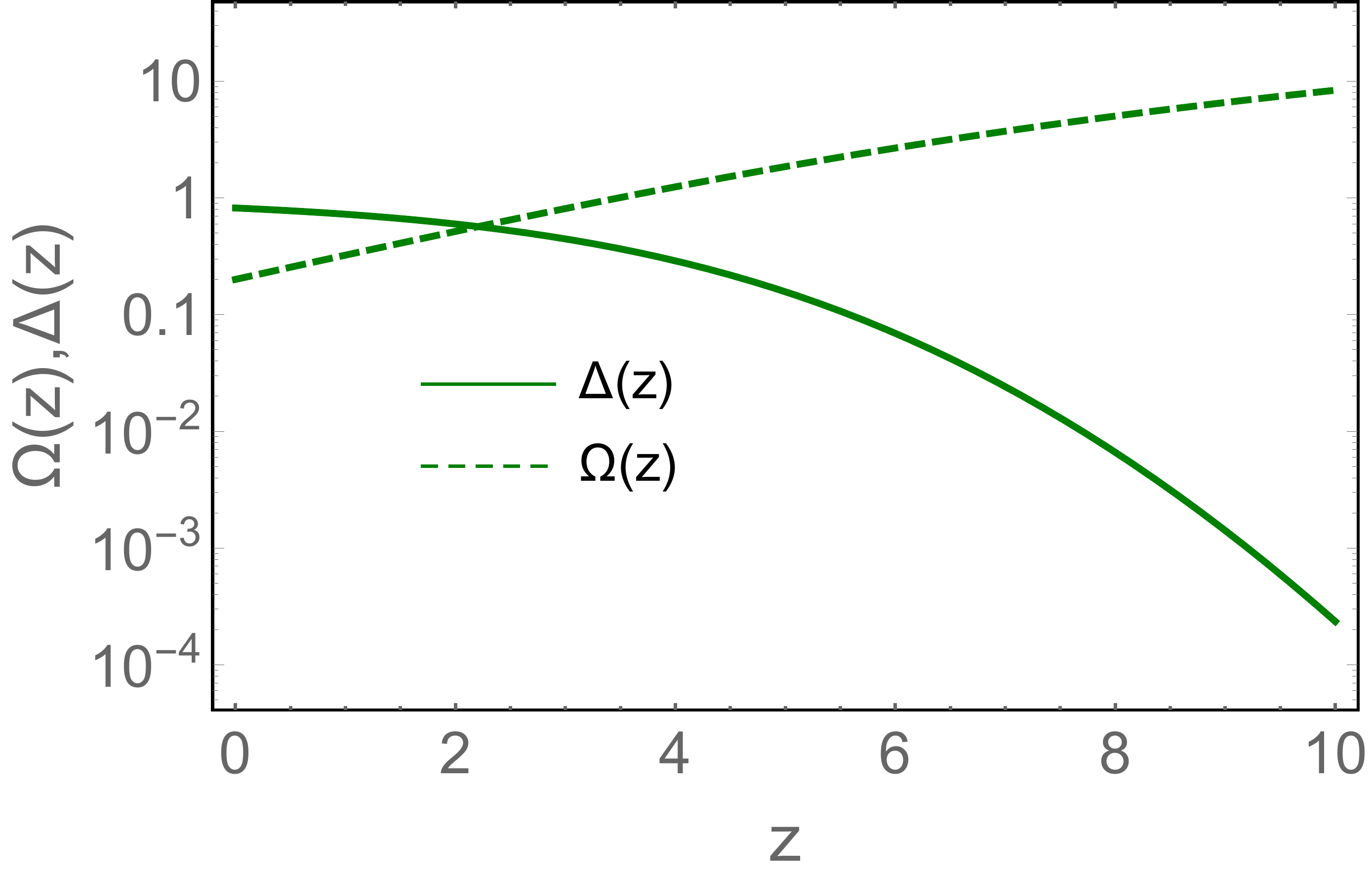} & & \includegraphics[width=6.8cm]{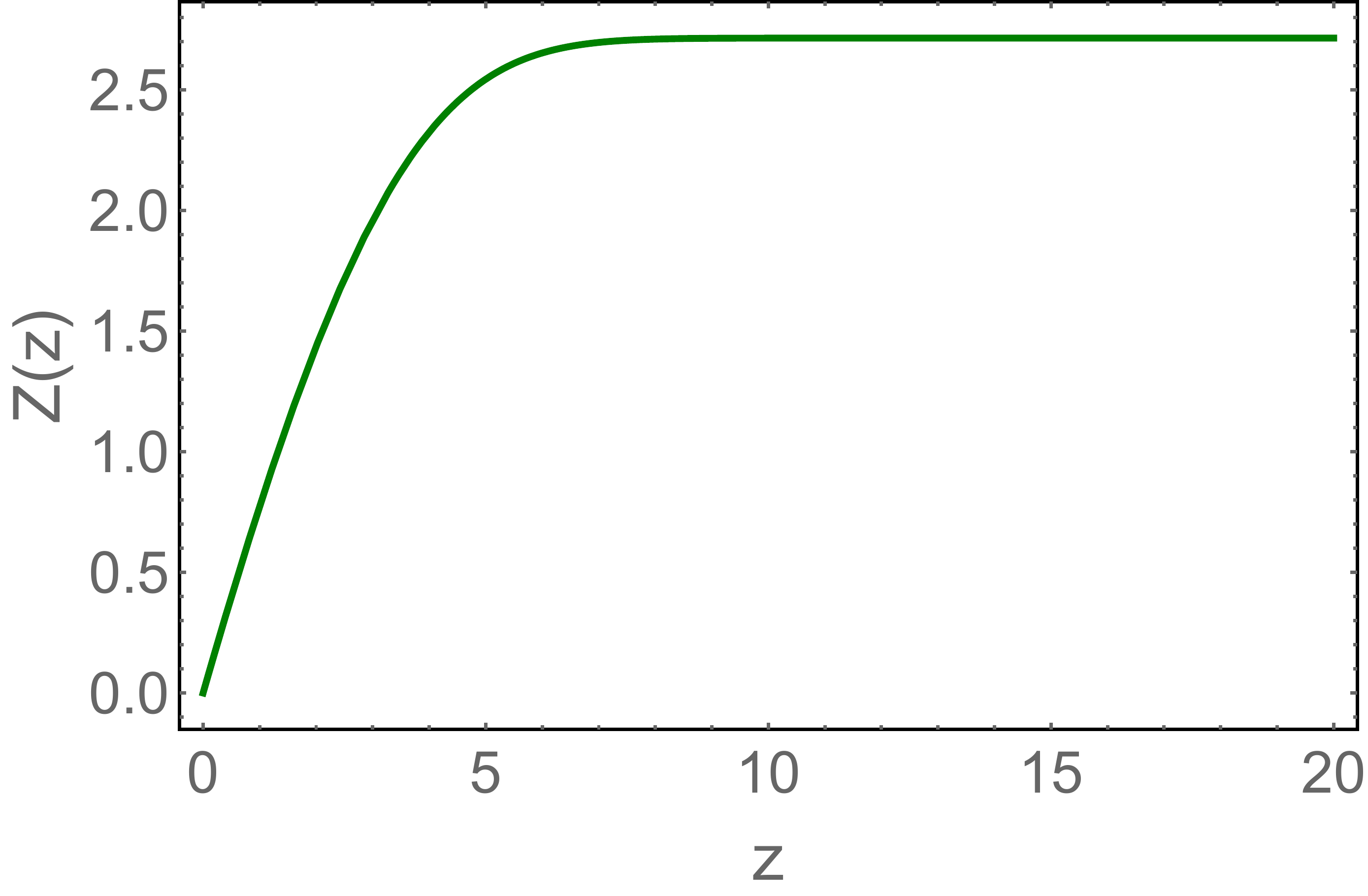}  \\
 \fig{omde}-a& ~~~~~ ~~~~~~~~~&\fig{omde}-b\\
 \end{tabular}
        \caption{ \fig{omde}-a: Functions $\Delta\Lb z\Rb$ and
 $\Omega\Lb z\Rb$ from \eq{SOL} for $\Omega_0(z=0) = 0.2$. 
 \fig{omde}-b: Function  $ Z(z)$ from \eq{ZZ} in which $\Omega(z)$ is
 calculated from \eq{SOL} with $\Omega_0 = 0.2$. }      
\label{omde}
\end{figure}

 As we have discussed in section II function $\Delta\Lb z\Rb$ depends on one
 variable $z\,\,=\,\,\ln\Lb r^2\,Q^2_s\Lb Y,b\Rb\Rb$, where $Q_s$ is the
 saturation momentum determined by the Baltsky-Kovchegov equation
 (see \eq{BK}). As one can see from \fig{omde}, the function $\Delta$ 
 decreases at large $z$ and is peaked at $z=0$.
 
 \subsection{Equation in  the leading twist approximation}


Using \eq{K2} for the BFKL kernel in the leading twist approximation, we
 we can reduce \eq{NEQ1} to the following equation:
 \bea \label{LT1}
&&\frac{\partial}{\partial Y}\,O\Lb \vec{x}_{10}, \vec{b} ,  Y; R \Rb \,\,=\,\,
-  \bas\!\! \int^{x^2_{01}}_{1/Q^2_s} \frac{d x^2_{02}}{x^2_{02}}\,\, O\Lb \vec{x}_{01},\vec{b}, Y; R\Rb\,\,\\
&&+\,\,\bas\,\Delta\Lb \vec{x}_{01},\vec{b} , Y; R\Rb\,\int^{x^2_{01}}_{1/Q^2_s} \frac{d x^2_{02}}{x^2_{02}}\,\,O\Lb \vec{x}_{02},\vec{b} , Y; R\Rb\,\,+\,\,\bas\,\int^{x^2_{01}}_{1/Q^2_s} \frac{d x^2_{02}}{x^2_{02}}\,\,\Delta\Lb \vec{x}_{02},\vec{b} , Y; R\Rb\,O\Lb \vec{x}_{01},\vec{b}, Y; R\Rb\nn\eea
 
Looking for the solution in the form
\beq \label{LT2}
O\Lb \vec{x}_{10}, \vec{b} ,  Y; R \Rb\,\,=\,\,o\Lb \vec{x}_{10}, \vec{b} ,  Y; R \Rb\,e^{ - \Omega\Lb z\Rb}
\eeq
we obtain the following equation for $o\Lb \vec{x}_{10}, \vec{b} ,  Y; R \Rb$:
\beq \label{LT3}
\frac{\partial}{\partial Y}\,o\Lb\xi, Y \Rb \,\,-\,\,\frac{d\, \Omega\Lb z\Rb}{ d Y}\,o\Lb \xi, Y \Rb=\,\,- \bas \,z\,o\Lb \xi, Y \Rb\,\,+\,\,\bas\,\,\int^{\xi} d \,\xi'\,\,o\Lb \xi', Y\Rb \,e^{ - \Omega\Lb z\Rb}\,\,+\,\,\bas\,\int^{z} d\,z'\,\,\Delta\Lb z'\Rb\,o\Lb \xi, Y\Rb
\eeq
where $\xi$ is defined by \eq{WW}. The first equation of \eq{BK4} can be
 rewritten in the form:
\beq \label{LT4}
\frac{\partial}{\partial Y}\,\Omega\Lb z \Rb\,\,=\,\,\bas\Lb\,z\, \,-\, \,\,\int^{z} d\,z'\,\,\Delta\Lb z'\Rb\Rb
\eeq
Plugging this equation into \eq{LT3} we reduce it to the form:
\beq \label{LT5}
\frac{\partial}{\partial Y}\,o\Lb\xi, Y \Rb\,\,\,=\,\,\,\bas\,\,\int^{\xi} d \,\xi'\,\,o\Lb \xi', Y\Rb \,e^{ - \Omega\Lb z'\Rb};~~~~~~~~~\frac{\partial^2}{\partial Y\,\partial \,\xi}\,o\Lb\xi, Y \Rb\,\,\,=\,\,\,\bas e^{- \,\Omega\Lb z\Rb}\,
o\Lb\xi, Y \Rb
\eeq
Introducing a new variable 
\beq \label{ZZ}
Z\,\,\,=\,\,\,\int^z_0\!\!\!d\,z' \,e^{ - \,\Omega\Lb z'\Rb}
\eeq
 we re-write the equation as follows:
\beq \label{LT6}
\frac{\partial^2}{\partial {Z}\,\partial\,\xi}\,o\Lb\xi, {Z} \Rb\,\,=\,\,\frac{1}{4}\,o\Lb\xi, {Z} \Rb
\eeq
\subsection{Solution}

The solution to \eq{LT6}  takes the general form:
\beq \label{LTSOL1}
o\Lb\xi, {Z} \Rb\,\,\,=\,\,\,\int^{\epsilon\,+\,i\,\infty}_{\epsilon\,-\,i\,\infty}\frac{d \,\gamma}{2\,\pi\,i}e^{\frac{1}{4}\frac{1}{\gamma}\,Z\,\,+\,\,\gamma\xi}\,\tilde{o}\Lb \gamma\Rb
\eeq
where $\tilde{o}\Lb \gamma\Rb$ should be found from the initial
 condition at $z=0$($Z=0$). We need to solve \eq{DLA4} to find
 this initial condition. Its solution has the form:
\beq \label{LTSOL2}
O^{\rm DLA}\Lb \xi'; Y \Rb\,\,=\,\,\frac{w}{w^*}\,O_0 \int^{\epsilon\,+\,i\,\infty}_{\epsilon\,-\,i\,\infty}\frac{d \,\gamma}{2\,\pi\,i}\,e^{\frac{\bas}{\gamma}\,Y\,\,+\,\,\gamma\xi' - \frac{3}{2} \xi'}\,\frac{1}{\gamma}
\eeq
with the initial condition $O^{\rm DLA}\Lb \xi'; Y=0 \Rb\,\,=
\,\,\frac{w}{w^*}\,\Lb w\,w^*\Rb^{3/2}\,O_0$, where $O_0$ is a
 constant. On the line $z=0$ \eq{LTSOL2} gives
\beq \label{LTSOL3}
O^{\rm DLA}\Lb \xi'; z=0 \Lb4 \bas Y = \xi' \Rb\Rb\,\,=\,\,\,O_0 \int^{\epsilon\,+\,i\,\infty}_{\epsilon\,-\,i\,\infty}\frac{d \,\gamma}{2\,\pi\,i}\,e^{\frac{1}{4}\frac{1}{\gamma}\,\xi'\,\,+\,\,\gamma\xi' - \frac{3}{2} \xi'}\,\frac{1}{\gamma}
\eeq
Taking into account \eq{LT2} we obtain the following initial condition
 for $o\Lb\xi, {Z} \Rb$:
\beq \label{LTSOL4}
o\Lb\xi, Z=0 \Rb\,\,\,=\,\,\,\int^{\epsilon\,+\,i\,\infty}_{\epsilon\,-\,i\,\infty}\frac{d \,\gamma}{2\,\pi\,i}e^{\gamma\xi}\,\tilde{o}\Lb \gamma\Rb\,\,=\,\,\,O_0\,e^{\Omega_0}\int^{\epsilon\,+\,i\,\infty}_{\epsilon\,-\,i\,\infty}\frac{d \,\gamma}{2\,\pi\,i}\,e^{\frac{1}{4}\frac{1}{\gamma}\,\xi'\,\,+\,\,\gamma\xi' - \frac{3}{2} \xi'}\,\frac{1}{\gamma}
\eeq

The r.h.s. of this equation we can re-write as 
\beq \label{LTSOL41}
O_0\,e^{\Omega_0}\int^{\epsilon\,+\,i\,\infty}_{\epsilon\,-\,i\,\infty}\frac{d \,\gamma'}{2\,\pi\,i}\,\frac{d \gamma(\gamma')}{d \,\gamma'}\,e^{\gamma' \xi}\,\frac{1}{\gamma(\gamma')}
\eeq
where $\gamma(\gamma')$ is the solution top the following equation:

\beq \label{LTSOL42}
\frac{1}{4}\frac{1}{\gamma}\,\,\,+\,\,\gamma - \frac{3}{2} = -\, \gamma'
\eeq

The solution to this equation gives:

\beq  \label{LTSOL43}
\gamma_{\pm}(\gamma')\,\,=\,\,\frac{1}{2} \Bigg(\frac{3}{2} \,-\,\gamma'\, \pm \,\sqrt{\Lb \gamma' - \frac{1}{2}\Rb\,\Lb \gamma' - \frac{5}{2}\Rb}\Bigg)
\eeq
Plugging \eq{LTSOL43} into \eq{LTSOL41} we obtain

\beq \label{LTSOL5}
\tilde{o}\Lb \gamma\Rb\,\,=\,\,\,O_0\,e^{\Omega_0}\,\frac{1}{\sqrt{\Lb \gamma - \h\Rb\,\Lb \gamma - \frac{5}{2}\Rb}}
\eeq
Hence the solution takes the form:
\beq \label{LTSOL6}
o\Lb\xi, {Z} \Rb\,\,\,=\,\,\,\,O_0\,e^{\Omega_0}\,\int^{\epsilon\,+\,i\,\infty}_{\epsilon\,-\,i\,\infty}\frac{d \,\gamma}{2\,\pi\,i}e^{\frac{1}{4}\frac{1}{\gamma}\,Z\,\,+\,\,\gamma\xi}\,\frac{1}{\sqrt{\Lb \gamma - \h\Rb\,\Lb \gamma - \frac{5}{2}\Rb}}
\eeq
For large $z$ and $\xi$, we can evaluate this integral using the method of
 steepest descend. For the saddle point we have the following equation:

\beq \label{LTSOL7}
-\frac{Z}{4 \,\gamma^2_{\rm sp}}\,\,+\,\,\xi\,\,=\,\,0;~~~~~\gamma_{\rm sp}\,\,=\,\,\h \sqrt{\frac{Z}{\xi}}\,\,\ll\,1 \,\,\mbox{for}\,\,\,\xi\,\,\gg\,\,1
\eeq

From \eq{LTSOL7} we obtain the solution:\,\,
\beq \label{LTSOL8}
o\Lb\xi, Z \Rb\,\,\,=\,\,\,\,O_0\,e^{\Omega_0}\,\sqrt{\pi}\frac{\gamma^3_{\rm sp}}{4\,Z}\frac{2}{ \sqrt{5}}\,e^{\sqrt{Z\,\xi}}
\eeq
which leads to the Odderon contribution
\beq \label{LTSOL9}
O\Lb\xi, Z \Rb\,\,\,=\,\,\,\frac{w}{w^*}\,O_0\,e^{\Omega_0\,-\,\Omega\Lb z\Rb}\,\sqrt{\pi}\frac{\gamma^3_{\rm sp}}{2\,Z}\,e^{\sqrt{Z\,\xi}}
\eeq

\begin{figure}
\centering
      \includegraphics[width=8cm]{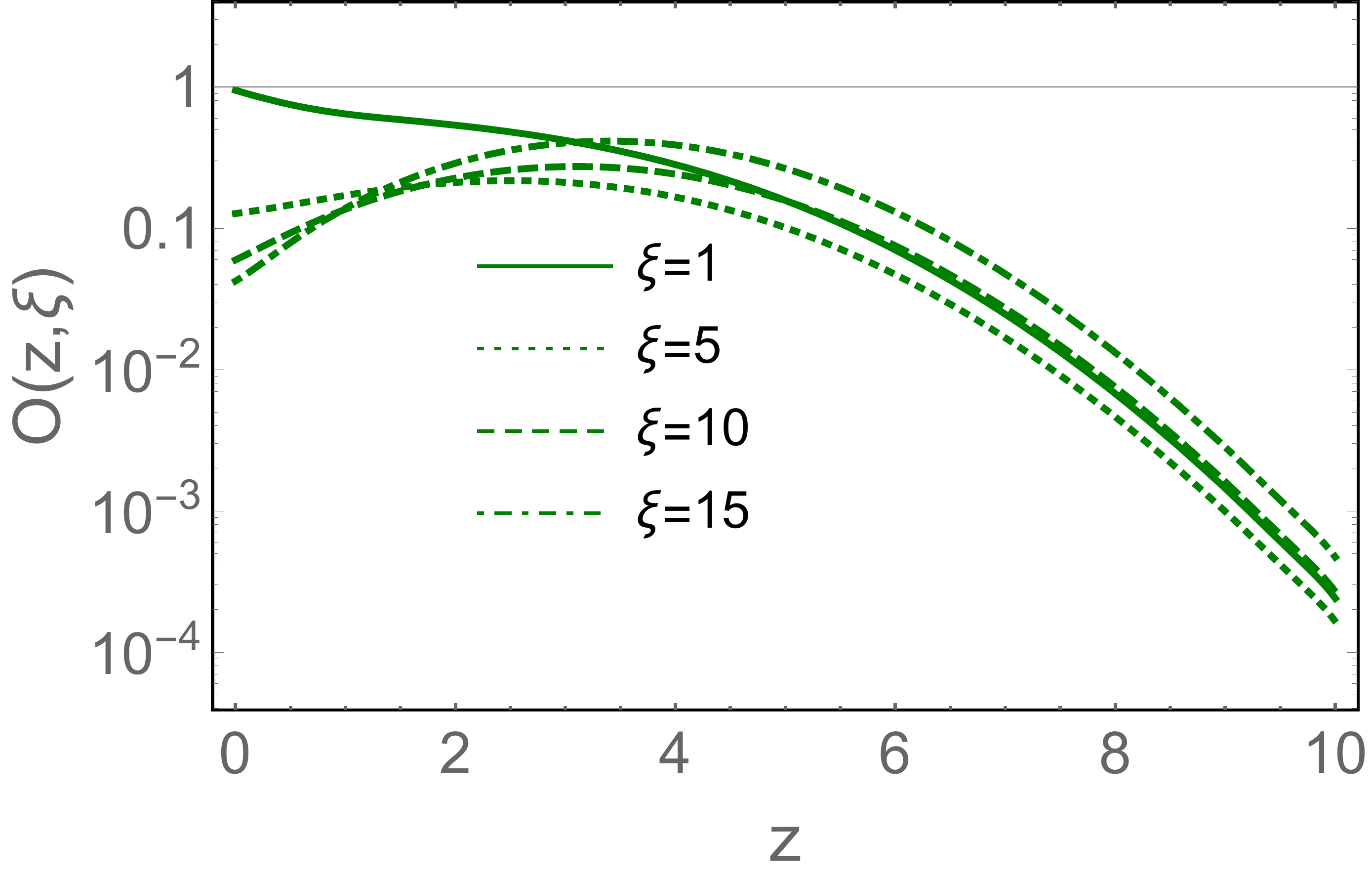} 
         \caption{ Solution for $O\Lb\xi, Z \Rb$ of \eq{LTSOL9} 
 versus $z$ at fixed $\xi$. $\Omega_0$=0.2, $O_0$=0.2. }      
\label{oo}
\end{figure}

In \fig{oo} solutions of \eq{LTSOL9} are plotted as a function of $z$ at
 fixed $\xi$.
One can see that all solutions lead to the Odderon contribution which
 is negligibly small at $z \,\geq 5$.

\section{Conclusions}
In the paper we proposed and solved  analytically the non-linear evolution equation in
 the leading twist approximation for the Odderon contribution. We found
 three qualitative features of this solution, where  the Odderon
 contribution differs from the Pomeron one:  (i )  the behaviour in the vicinity of the saturation scale cannot be  derived from the linear evolution in  a dramatic difference with the Pomeron case; (ii) the substantial
 decrease of the Odderon contribution with the energy; and (iii)  
  the lack of the geometric scaling behaviour.  All of these features
 can be see from \fig{oo} and \eq{LTSOL9}.
The decrease of the Odderon contribution with energy should be
 confronted with the QCD linear equation prediction for the
 intercept of the Odderon:  $\alpha_{\rm Odd}\Lb t=0\Rb\,=\,1$, which
 means that the Odderon contribution does not depend on energy. The
 geometric scaling behaviour is the most striking general feature
 of the non-linear Balitsky-Kovchegov equation\cite{LETU,BALE,IIML,SGBK}.
 Therefore, the Odderon provides an example that this behaviour is the
 characteristic property of the Pomeron contributions. It is instructive
 to mention, that in spite of the violation of the geometric scaling
 behaviour for the Odderon, the suppression deep in the saturation
 region is the same as for the Pomeron case, and is determined by
 the  gluon reggeization \cite{LETU}. We would like to stress that some of these features( the decrease with energy and the lack of the geometric scaling behaviour)  have been seen in the numerical solutions  to the non-linear evolution of the QCD Odderon\cite{LRRW,YHH}; and the decrease of the Odderon contribution with energy  follows from the general non-linear equation\cite{KS,KOLEB}, using the approach of Ref.\cite{LETU}.

Concluding we wish to stress that the QCD Odderon gives a very small
 contribution to the scattering amplitude, due to substantial
 shadowing corrections, which are responsible for the non-linear
 evolution. We believe that this solid theoretical result based
 on the effective QCD theory  at high energy: the CGC approach, will
 be useful in our discussion of the available experimental data.

\section{Acknowledgements}
   We thank our colleagues at Tel Aviv university and UTFSM for
 encouraging discussions. Our special thanks go Errol Gotsman for his everyday advices and discussions on the subject of the paper. We wish to thank Yoshitaka Hatta, who drew our attention to the numerical solutions for the QCD Odderon, that we overlooked. 
 
   This research was supported  by 
   Proyecto Basal FB 0821(Chile),  Fondecyt (Chile) grants  
 1180118 and 1191434.  
  
\end{document}